\documentclass[runningheads]{llncs}

\usepackage[T1]{fontenc}
\usepackage{amsmath,amssymb,amsfonts}
\usepackage{xspace}
\usepackage{booktabs}
\usepackage{color}
\usepackage{xcolor}
\usepackage{graphicx}
\usepackage{tikz}
\usetikzlibrary{arrows,decorations.pathmorphing,decorations.pathreplacing,positioning,fit,trees,shapes,shadows,automata,calc} 
\usepackage{wrapfig}
\usepackage{setspace}
\usepackage{hyperref}
\usepackage{pgfplots}
\usepackage{subfigure}

\usepackage{marginnote}

\usepackage{cleveref}
\crefformat{footnote}{#2\footnotemark[#1]#3}

\newcommand{\tool}[1]{\textsc{#1}\xspace}
\newcommand{\prism}{\tool{Prism}}
\newcommand{\storm}{\tool{Storm}}
\newcommand{\mrmc}{\tool{MRMC}}
\newcommand{\iscasmc}{\tool{iscasMC}}
\newcommand{\imca}{\tool{IMCA}}
\newcommand{\param}{\tool{PARAM}}
\newcommand{\epmc}{\tool{Epmc}}
\newcommand{\engine}[1]{\emph{#1}\xspace}
\newcommand{\maxsat}{\textsc{MaxSat}\xspace}

\selectcolormodel{cmy}

\renewcommand{\paragraph}[1]{\par\smallskip\noindent\emph{#1}}
\renewcommand{\subsubsection}[1]{\par\smallskip\noindent\textbf{#1}}

\begin{document}
  \title{A \storm is Coming: \\ A Modern Probabilistic Model Checker}
\author{Christian Dehnert \and Sebastian Junges \and Joost-Pieter Katoen \and Matthias Volk}
\institute{RWTH Aachen University, Aachen, Germany}
\titlerunning{A \storm is Coming: A Modern Probabilistic Model Checker}
\authorrunning{C. Dehnert et al.}
\maketitle
  \begin{abstract}
We launch the new probabilistic model checker \storm{}.
It features the analysis of discrete- and continuous-time variants of both Markov chains and MDPs.
It supports the \prism{} and JANI modeling languages, probabilistic programs, dynamic fault trees and generalized stochastic Petri nets. 
It has a modular set-up in which solvers and symbolic engines can easily be exchanged.
It offers a Python API for rapid prototyping by encapsulating \storm's fast and scalable algorithms. 
Experiments on a variety of benchmarks show its competitive performance.
\end{abstract}
  \section{Introduction}
In the last five years, we have developed our in-house probabilistic model checker with the aim to have an easy-to-use platform for experimenting with new verification algorithms, richer probabilistic models, algorithmic improvements, different modeling formalism, various new features, and so forth.
Although open-source probabilistic model checkers do exist, most are not flexible and modular enough to easily support this.
Our efforts have led to a toolkit with mature building bricks with simple interfaces for possible extensions, and a modular set-up.
It comprises about 100,000 lines of \texttt{C++} code.
The time has come to make this toolkit available to a wider audience: this paper presents \storm.

Like its main competitors \prism~\cite{KNP11}, \mrmc~\cite{mrmc}, and \iscasmc~\cite{iscasmc}, \storm{} relies on numerical and symbolic computations.
It does not support discrete-event simulation, known as statistical model checking~\cite{DBLP:conf/isola/LarsenL16}.
The main characteristic features of \storm{} are:
\begin{itemize}
\item
it supports \emph{various native input formats}: the \prism{} input format, generalized stochastic Petri nets, dynamic fault trees, and conditioned probabilistic programs. 
This is not just providing another parser; state-space reduction and generation techniques as well as analysis algorithms are partly tailored to these modeling formalisms;
\item 
in addition to Markov chains and MDPs, it supports \emph{Markov automata}~\cite{DBLP:conf/lics/EisentrautHZ10,DBLP:journals/iandc/DengH13}, a model containing probabilistic branching, non-determinism, and exponentially distributed delays;
\item
it can do \emph{explicit state} and \emph{fully symbolic} (BDD-based) model checking as well as a \emph{mixture} of these modes;
\item 
it has a \emph{modular} set-up, enabling the easy exchange of different solvers and distinct decision diagram packages; its current release supports about 15 solvers, and the BDD packages {\tt CUDD}~\cite{cudd_website} and multi-threaded {\tt Sylvan}~\cite{DBLP:conf/tacas/DijkP15};
\item
it provides a \emph{Python API} facilitating easy and rapid prototyping of other tools using the engines and algorithms in \storm;
\item
it provides the following functionalities under one roof: the synthesis of counterexamples and permissive schedulers (both MILP- and SMT-based), game-based abstraction of infinite-state MDPs, efficient algorithms for conditional probabilities and rewards~\cite{baier_cond}, and long-run averages on MDPs~\cite{DBLP:conf/lics/Alfaro98};
\item
its performance in terms of verification speed and memory footprint on the \prism benchmark suite is mostly better compared to \prism{}.
\end{itemize}
Although many functionalities of \prism{} are covered by \storm, there are significant differences.
\storm{} does not support LTL model checking (as in \iscasmc{} and \prism) and does not support the \prism{} features: probabilistic timed automata, multi-objective model checking, and an equivalent of \prism's ``hybrid'' engine (a crossover between full MTBDD and \storm's ``hybrid'' engine), a fully symbolic engine for continuous-time models, statistical model checking, and the analysis of stochastic games as in \prism-\tool{GAMES}~\cite{DBLP:conf/tacas/ChenFKPS13}.

  \section{Features}
\paragraph{Model types.} 
\storm{} supports Markov chains and Markov decision processes (MDPs), both in two forms: discrete time and continuous time.
This yields four different models: classical discrete-time (DTMCs) and continuous-time Markov chains (CTMCs), as well as MDPs and Markov automata (MA)~\cite{DBLP:conf/lics/EisentrautHZ10,DBLP:journals/iandc/DengH13}, a compositional variant of continuous-time MDPs.
The MA is the richest model.
CTMCs are MAs without non-determinism, while MDPs are MAs without delays; DTMCs are CTMCs without delays, cf.~\cite{Kat16}.
All these models are extensible with rewards (or dually: costs) to states, and -- for non-deterministic models -- to actions.
Most probabilistic model checkers support Markov chains and/or MDPs; MAs so far have only been supported by few tools~\cite{DBLP:conf/atva/GuckTHRS14,DBLP:journals/corr/GuckHHKT14}.
\paragraph{Modeling languages.}
\storm{} supports various symbolic and an explicit input format to specify the aforementioned model types: 
(i) Most prominently, the \prism{} input language~\cite{prism_website};
(ii) the recently defined JANI format~\cite{jani}, a universal  probabilistic modeling language;
(iii) as the first tool \emph{every}\footnote{Existing CSL model checkers for GSPNs such as GreatSPN~\cite{DBLP:conf/apn/AmparoreBD14} and MARCIE~\cite{DBLP:conf/qest/SchwarickHR11} are restricted to confusion-free Petri nets; \storm does not have this restriction as it supports MA.}
 generalized stochastic Petri net (GSPN)~\cite{DBLP:conf/apn/EisentrautHK013} via both a dedicated model builder as well as an encoding in JANI;
(iv) dynamic fault trees (DFTs)~\cite{Dugan1992,Boudali2010} --
due to dedicated state-space generation and reduction techniques for DFTs, \storm{} significantly outperforms competing tools in this domain~\cite{DBLP:conf/safecomp/0001JK16};
(v) pGCL probabilistic programs~\cite{DBLP:series/mcs/McIverM05} extended with observe-statements~\cite{DBLP:conf/icse/GordonHNR14}, an essential feature to describe and analyze e.g., Bayesian networks;
(vi) in the spirit of MRMC~\cite{mrmc}, models can be provided in a format that explicitly enumerates transitions. 
\paragraph{Properties.}
\storm{} focusses on probabilistic branching-time logics, i.e.~PCTL \cite{HJ94} and CSL \cite{DBLP:conf/cav/AzizSSB96,DBLP:journals/tse/BaierHHK03} for discrete-time and continuous-time models, respectively. 
To enable the treatment of reward objectives such as expected and long-run rewards, \storm{} supports reward extensions of these logics in a similar way as \prism{}.
In addition, \storm{} supports conditional probabilities and conditional rewards~\cite{baier_cond}; these are, e.g., important for the analysis of cpGCL programs. 
\paragraph{Engines.}
\storm{} features two distinct in-memory representations of probabilistic models:  \engine{sparse matrices} allow for fast operations on small and moderately sized models, multi-terminal binary decision diagrams (MTBDDs) are able to represent gigantic models, however with slightly more expensive operations.
A variety of engines built around the in-memory representations is available, which allows for the more efficient treatment of input models.
Both \storm's \engine{sparse} and the \engine{exploration} engine purely use a sparse matrix-based representation. While the former amounts to an efficient implementation of the standard approaches, the latter one implements the ideas of \cite{DBLP:conf/atva/BrazdilCCFKKPU14} which scrutinizes the state space with machine learning methods.
Three other engines, \engine{dd}, \engine{hybrid} and \engine{abstraction-refinement}, use MTBDDs as their primary representation.
While \engine{dd} exclusively uses decision diagrams, \engine{hybrid} also uses sparse matrices for operations deemed more suitable on this format.
The \engine{abstraction-refinement} engine abstracts (possibly infinite) discrete-time Markov models to (finite) stochastic games and automatically refines the abstraction as necessary.
\paragraph{Parametric models.}
\storm was used as backend in \cite{dehnert-et-al-cav-2015,DBLP:conf/atva/QuatmannD0JK16}. By using the dedicated library \tool{CArL} \cite{carl_website} for the representation of rational functions and applying novel algorithms for the analysis of parametric discrete-time models, it has proven to significantly outperform other tools such as the parametric algorithms in \prism{} and the dedicated tool \param~\cite{param_sttt}. 
\paragraph{Exact arithmetic.}
Several works \cite{DBLP:conf/rp/HaddadM14,DBLP:conf/mmb/WimmerB10} observed that the numerical methods applied by probabilistic model checkers are prone to numerical problems. \storm therefore supports enabling exact arithmetic to obtain \emph{precise results}.
\paragraph{Counterexample generation.}
For probabilistic models, several counterexample representations have been proposed~\cite{DBLP:conf/sfm/AbrahamBDJKW14,DBLP:conf/cav/BrazdilCCFK15}. 
\storm implements the MILP-based counterexample technique~\cite{DBLP:journals/corr/abs-1305-5055}, as well as the \maxsat-based generation of high-level counterexamples on \prism{} models~\cite{DBLP:conf/atva/DehnertJWAK14}.
These algorithms go beyond the capabilities of dedicated, stand-alone counterexample generation tools such as DiPro~\cite{ALLS11} and COMICS~\cite{jansen-et-al-atva-2012}.
In particular, the synthesis of high-level counterexamples facilitates to obtain counterexamples as \prism{} code, starting from a \prism{} model and a refuted property.
\paragraph{APIs.}
\storm{} can be used via \emph{three} interfaces: a command-line interface, a C++ API, and a Python API. 
The command-line interface consists of several binaries that provide end-users access to the available settings for different tasks.  
Advanced users can utilize the many settings to tune the performance.
Developers may either use the C++ API that offers fine-grained and performance-oriented access to \storm's functionality, or the Python API which allows rapid prototyping and encapsulates the high-performance implementations within \storm.

  \section{Architecture}
\begin{figure}[t]
  \begin{center}
  \scalebox{0.8}{
    \begin{tikzpicture}[node distance=0.7cm]
      \tikzstyle{ast}=[text width=1cm, align=right, minimum height=0.5cm]
      \tikzstyle{builder}=[minimum width=1.5cm, minimum height=0.7cm]
      \tikzstyle{builderinput}=[->, shorten <=5pt, shorten >=5pt]
      \tikzstyle{builderoutput}=[]
      \tikzstyle{feature}=[text width=1.5cm, align=center, minimum height=0.5cm]
      \tikzstyle{solver}=[minimum height=0.5cm, text width=1.2cm, align=center]
      \tikzstyle{solveruse}=[dashed]
    
      \node[ast] (ast_cpgcl) {cpGCL};
      \node[ast, below of=ast_cpgcl] (ast_dft) {DFT};
      \node[ast, below of=ast_dft] (ast_gspn) {GSPN};
      \node[ast, below of=ast_gspn] (ast_prism) {PRISM};
      \node[ast, below of=ast_prism] (ast_jani) {JANI};
      
      \node[right=1cm of ast_dft, builder] (builder_dft) {};
      \node[right=1cm of ast_gspn, builder] (builder_gspn) {};
      \node[right=1cm of ast_prism, builder] (builder_prism) {};
      \node[right=1cm of ast_jani, builder] (builder_jani) {};

      \coordinate (builder_bottom_left) at ($(ast_jani.south east)+(1cm,0)+0.5*(ast_jani.north)-0.5*(ast_prism.south)$);
      \coordinate (builder_top_left) at ($(ast_dft.north east)+(1cm,0)-0.5*(ast_jani.north)+0.5*(ast_prism.south)$);
      
      \draw[rounded corners] ($(builder_top_left)+(0,0.9cm)$) rectangle ($(builder_bottom_left)+(1.5,0)$);
      \node[above=0cm of builder_dft, text width=2cm, align=center] (builder) {model \\ builder};
      
      \draw (builder_dft.north west) edge[thick, dotted] (builder_dft.north east);
      \draw (builder_gspn.north west) edge[thick, dotted] (builder_gspn.north east);
      \draw (builder_prism.north west) edge[thick, dotted] (builder_prism.north east);
      \draw (builder_jani.north west) edge[thick, dotted] (builder_jani.north east);
      
      \coordinate (block1_left) at (builder_dft.west);
      \coordinate (block2_left) at (builder_gspn.west);
      \coordinate (block3_left) at (builder_prism.west);
      \coordinate (block4_left) at (builder_jani.west);

      \coordinate (block1_right) at (builder_dft.east);
      \coordinate (block2_right) at (builder_gspn.east);
      \coordinate (block3_right) at (builder_prism.east);
      \coordinate (block4_right) at (builder_jani.east);
      
      \node[feature, right=1.8cm of block1_right] (feat_counterexamples) {counterex.};
      \node[feature, right=1.8cm of block2_right] (feat_sparse) {sparse};
      \node[feature, above=0.2cm of feat_counterexamples] (feat_permsched) {permissive \\ schedulers};

      \node[feature, right=1.8cm of block3_right] (feat_hybrid) {hybrid};
      \node[feature, right=1.8cm of block4_right] (feat_dd) {dd};
      \node[feature, below=0.2cm of feat_dd] (feat_exploration) {exploration};
      \node[feature, below=0.2cm of feat_exploration] (feat_absref) {abstr.-ref.};

      \coordinate (t1) at (feat_permsched.east);
      \coordinate (t2) at (feat_absref.east);
      \draw let \p1 = (t1), \p2 = (t2), \n1 = {0*veclen((\x2-\x1),(\y2-\y1))}
      in
      node[solver, anchor=west] (solver_smt) at ($(feat_permsched.east)+(1.7,-\n1)$) {SMT};
      
      \draw let \p1 = (t1), \p2 = (t2), \n1 = {0.25*veclen((\x2-\x1),(\y2-\y1))}
      in
      node[solver, anchor=west] (solver_milp) at ($(feat_permsched.east)+(1.7,-\n1)$) {(MI)LP};

      \draw let \p1 = (t1), \p2 = (t2), \n1 = {0.5*veclen((\x2-\x1),(\y2-\y1))}
      in
      node[solver, anchor=west] (solver_linear) at ($(feat_permsched.east)+(1.7,-\n1)$) {linear};

      \draw let \p1 = (t1), \p2 = (t2), \n1 = {0.75*veclen((\x2-\x1),(\y2-\y1))}
      in
      node[solver, anchor=west] (solver_bellman) at ($(feat_permsched.east)+(1.7,-\n1)$) {Bellman};

      \draw let \p1 = (t1), \p2 = (t2), \n1 = {1*veclen((\x2-\x1),(\y2-\y1))}
      in
      node[solver, anchor=west] (solver_games) at ($(feat_permsched.east)+(1.7,-\n1)$) {games};
            
      \draw (ast_dft.east) edge[builderinput] (block1_left);
      \draw (ast_gspn.east) edge[builderinput] (block2_left);
      \draw (ast_prism.east) edge[builderinput] (block3_left);
      \draw (ast_jani.east) edge[builderinput] node [pos=0.5,inner sep=0pt,minimum size=0pt] (jani_split) {} (block4_left);

      \coordinate (brace_offset) at (0.3,0);
      \coordinate (block1_right_offset) at ($(block1_right)+(brace_offset)$);
      \coordinate (block4_right_offset) at ($(block4_right)+(brace_offset)$);
      
      \coordinate (brace_offset2) at (0.15,0);
      \draw [decorate,decoration={brace,amplitude=5}] ($(block1_right)+(brace_offset2)$) -- node[inner sep=0pt, minimum size=9pt] (brace1_middle) {} ($(block4_right)+(brace_offset2)$);
      
      \coordinate (brace2_top) at ($(feat_permsched.west)-(brace_offset)$);
      \coordinate (brace2_bottom) at ($(feat_sparse.west)-(brace_offset)$);
      \coordinate (brace2_middle) at ($(brace2_top)!(feat_counterexamples.west)!(brace2_bottom)$);
      \draw[rounded corners, ->] (brace2_middle) -- (brace2_top) -- (feat_permsched.west);
      \draw[rounded corners, ->] (brace2_middle) -- (brace2_bottom) -- (feat_sparse.west);
      
      \coordinate (brace1_middle_feat_projection) at ($(brace2_top)!(brace1_middle)!(brace2_bottom)$);
      \coordinate (brace1_middle_out) at ($(brace1_middle)!0.5!(brace1_middle_feat_projection)$);
      \draw[rounded corners,->] (brace1_middle) -- (brace1_middle_out) -- (brace1_middle_out |- feat_counterexamples.west) -- (feat_counterexamples.west);
      
      
      \coordinate (block3_sec_right) at ($(block3_right)+(0.45,0)$);
      \coordinate (block4_sec_right) at ($(block4_right)+(0.45,0)$);
      \coordinate (block3_sec_right_offset) at ($(block3_sec_right)+(brace_offset)$);
      \coordinate (block4_sec_right_offset) at ($(block4_sec_right)+(brace_offset)$);

      \draw [decorate,decoration={brace,amplitude=5pt}] (block3_sec_right) -- node[inner sep=0pt, minimum size=9pt] (blocks34_sec_offset_middle) {} (block4_sec_right);

      \coordinate (feat_hybrid_sec_offset) at ($(feat_hybrid.west)-(brace_offset)$);
      \coordinate (feat_dd_sec_offset) at ($(feat_dd.west)-(brace_offset)$);
      \coordinate (hybrid_dd_sec_offset_middle) at ($(feat_hybrid_sec_offset)!0.5!(feat_dd_sec_offset)$);
      \draw (blocks34_sec_offset_middle) edge[builderoutput] (hybrid_dd_sec_offset_middle);
            
      \draw[rounded corners, ->] (hybrid_dd_sec_offset_middle) -- (feat_hybrid_sec_offset) -- (feat_hybrid.west);
      \draw[rounded corners, ->] (hybrid_dd_sec_offset_middle) -- (feat_dd_sec_offset) -- (feat_dd.west);
      
      
      \coordinate (feat_exploration_third_offset) at ($(feat_exploration.west)-(brace_offset)$);
      \coordinate (feat_absref_third_offset) at ($(feat_absref.west)-(brace_offset)$);
      \coordinate (brace5_middle) at ($(feat_exploration_third_offset)!0.5!(feat_absref_third_offset)$);
      \draw[rounded corners,->] (brace5_middle) -- (feat_exploration_third_offset) -- (feat_exploration.west);
      \draw[rounded corners,->] (brace5_middle) -- (feat_absref_third_offset) -- (feat_absref.west);
      
      \draw[rounded corners] (jani_split) -- (jani_split |- brace5_middle) -- (brace5_middle);
      
                      
      \coordinate (above_builder) at ($(builder_dft.north)+(0,1.1cm)$);
      \coordinate (above_smt) at ($(solver_smt.north)!(above_builder)!(solver_milp.north)$);
      \draw[rounded corners, dashed] ($(builder_dft.north)+(0,0.9cm)$) -- (above_builder) -- (above_smt) -- (solver_smt.north);
                      
      \draw (feat_absref.east) edge[solveruse] (solver_games.west);
      \draw[solveruse, rounded corners] (feat_absref.east) -- (solver_smt.west);

      \draw (feat_sparse.east) edge[solveruse] (solver_milp.west);
      \draw (feat_sparse.east) edge[solveruse] (solver_linear.west);
      \draw (feat_sparse.east) edge[solveruse] (solver_bellman.west);

      \draw (feat_counterexamples.east) edge[solveruse] (solver_smt.west);
      \draw (feat_counterexamples.east) edge[solveruse] (solver_milp.west);

      \draw (feat_permsched.east) edge[solveruse] (solver_smt.west);
      \draw (feat_permsched.east) edge[solveruse] (solver_milp.west);

      \draw (feat_hybrid.east) edge[solveruse] (solver_milp.west);
      \draw (feat_hybrid.east) edge[solveruse] (solver_bellman.west);
      \draw (feat_hybrid.east) edge[solveruse] (solver_linear.west);

      \draw (feat_dd.east) edge[solveruse] (solver_bellman.west);
      \draw (feat_dd.east) edge[solveruse] (solver_linear.west);

      \draw [decorate,decoration={brace,amplitude=5pt}] ($(solver_smt.north east)+(0.05,0)$) -- node[rotate=-90, yshift=0.35cm] (solver_text) {solver} ($(solver_games.south east)+(0.05,0)$);

      
      \coordinate (ast_cpgcl_offset) at ($(ast_cpgcl.west)-(brace_offset)$);
      \coordinate (ast_cpgcl_offset) at ($(ast_cpgcl.west)-(brace_offset)$);

      \draw[rounded corners, ->] (ast_cpgcl.west) -- ($(ast_cpgcl.west)-(brace_offset)$) -- ($(ast_jani.west)-(brace_offset)$) -- (ast_jani.west);

      \draw[rounded corners, ->] (ast_gspn.west) -- ($(ast_gspn.west)-(brace_offset)$) -- ($(ast_jani.west)-(brace_offset)$) -- (ast_jani.west);

      \draw[rounded corners, ->] (ast_prism.west) -- ($(ast_prism.west)-(brace_offset)$) -- ($(ast_jani.west)-(brace_offset)$) -- (ast_jani.west);

    \end{tikzpicture}
    }
  \end{center}
  \vspace{-3mm}
  \caption{\storm's architecture.}
  \label{fig:architecture}
\end{figure}
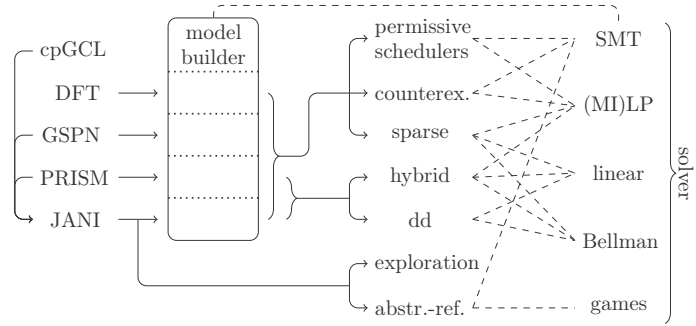
Fig.~\ref{fig:architecture} depicts the architecture of \storm. Solid arrows indicate the flow of control and data, dashed lines represent a ``uses'' relationship.
After the initial parsing step, it depends on the selected engine whether a model building step is performed: for all but the \engine{exploration} and \engine{abstraction-refinement} engines, it is necessary to build a full in-memory representation of the model upfront. Note that the available engines depend on the input format and that both PRISM and GSPN input can be either treated natively or transformed to JANI.
\paragraph{Solvers.}
\storm's infrastructure is built around the notion of a \emph{solver}.
For instance, solvers are available for sets of linear or Bellman equations (both using sparse matrices as well as MTBDDs), (mixed-integer) linear programming (MILP) and satisfiability modulo theories (SMT).
Note that \storm does not support stochastic games as input models, yet, but solvers for them are available because they are used in the \engine{abstraction-refinement} engine.
Offering these interfaces has several key advantages.
First, it provides easy and coherent access to the tasks commonly involved in probabilistic model checking.
Secondly, it enables the use of dedicated state-of-the-art high-performance libraries for the task at hand. More specifically, as the performance characteristics of different backend solvers can vary drastically for the same input, this permits choosing the best solver for a given task.
Licensing problems are avoided, because implementations can be easily enabled and disabled, depending on whether or not the particular license fits the requirements.
Finally, implementing new solver functionality is easy and can be done without knowledge about the global code base. It allows to embed new state-of-the-art solvers in the future.
For each of those interfaces, several actual implementations exist. Table~\ref{tab:solvers} gives an overview over the currently available implementations.
\begin{table}[t]
  \def\arraystretch{1.3}\tabcolsep=5pt
  \begin{center}
    \caption{Solvers offered by \storm.}
    \vspace{-2mm}
  \small{
  
    \begin{tabular}{ll}
      \toprule
      solver type & available solvers \\
      \cmidrule{1-2}
      linear equations (sparse) & Eigen \cite{eigenweb}, gmm++ \cite{gmmpp_website}, elim. \cite{Daw04}, built-in \\
      linear equations (MTBDD) & CUDD \cite{cudd_website}, Sylvan \cite{DBLP:conf/tacas/DijkP15} \\
      Bellman equations (sparse) & Eigen, gmm++, built-in \\
      Bellman equations (MTBDD) & CUDD, Sylvan \\
      stochastic games (sparse) & built-in \\
      stochastic games (MTBDD) & CUDD, Sylvan \\
      (MI)LP & Gurobi \cite{gurobi}, glpk \cite{glpkweb} \\
      SMT & Z3 \cite{dMB08}, MathSAT \cite{DBLP:conf/tacas/CimattiGSS13}, SMTLIB \cite{Barrett10c.:the} \\
      \bottomrule
    \end{tabular} 
    }
    \label{tab:solvers}
  \end{center}
  \vspace{-4mm}
\end{table}
Almost all engines and all other key modules make use of \emph{solvers}.
The most prominent example is the use of the equation solvers for answering standard verification queries. 
However, other modules use them too, e.g.\ model building (SMT), counterexample generation~\cite{DBLP:journals/corr/abs-1305-5055,DBLP:conf/atva/DehnertJWAK14} (SMT, MILP) and permissive scheduler generation~\cite{DBLP:journals/corr/DragerFK0U15,junges-et-al-tacas-2016} (SMT, MILP).

  \section{Evaluation}
\paragraph{Set-up.}
For the performance evaluation, we conducted experiments on a HP BL685C G7. Up to eight cores with 2.0GHz and 8GB of memory were available to the tools, but only \prism's garbage collection used more than one core at a time. We set a time-out of 1800 seconds.
\subsubsection{Comparison with \prism.}
To assess \storm's performance on standard model-checking queries, we compare it with \prism on the \prism benchmark suite \cite{KNP12b}. 
More specifically, we consider all DTMCs, CTMCs and MDPs (24 in total, and several instances per model) and all corresponding properties (82 in total). Note that we do not compare \storm with \iscasmc as the latter one has a strong focus on more complex LTL properties.\footnote{\label{note:appendix} More details and experiments can be found in Appendix \ref{app:moreexperiments}.}
\paragraph{Methodology.}
As both \prism and \storm offer several engines with different strengths and weaknesses, we choose the following comparison methodology. We compare engines that ``match'' in terms of the general approach.
For example, \prism's \engine{explicit} engine first builds the model in terms of a sparse matrix directly and then performs the model checking on this representation, which matches the approach of \storm's \engine{sparse} engine.
In the same manner, \prism's \engine{sparse} engine is comparable to \storm's \engine{hybrid} one and \prism's \engine{mtbdd} engine corresponds to \storm's \engine{dd} engine.
Finally, we compare the run-times of \prism and \storm when selecting the best engine for each individual benchmark instance.
\paragraph{Results.}
\newcommand{\yaxisspace}{0.23}
\newcommand{\xaxisspace}{-0.05}
\newcommand{\divider}{\hspace{0.7cm}}
\tikzset{%
    dtmc/.style={mark=square*,blue,mark size=0.9},
    mdp/.style={mark=triangle*,red,mark size=0.9},
    ctmc/.style={mark=*,brown, mark size=0.9},
    ma/.style={mark=diamond*,purple, mark size=0.9},
}
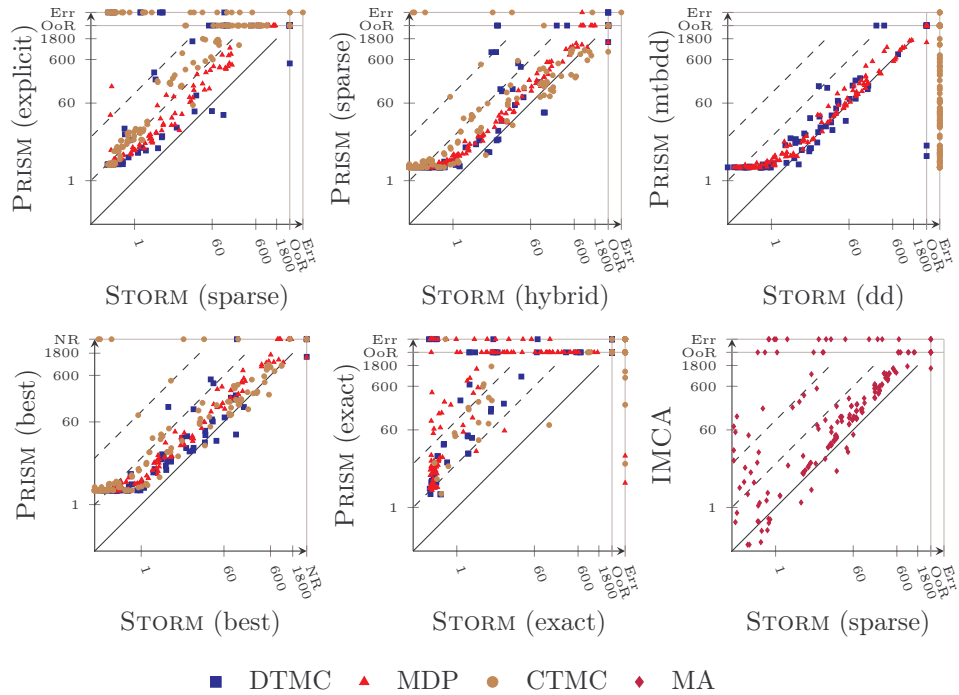
\begin{figure}[tb]
\hspace{-0.29cm}
    \begin{minipage}{0.28\textwidth}
      \centering
      \begin{tikzpicture}
  \begin{axis}[
    width=4.4cm, height=4.4cm, xmin=0.1, ymin=0.1, ymax=7200, xmax=7200, xmode=log, ymode=log,
    axis x line=bottom, axis y line=left,
    x label style={at={(axis description cs:0.5,\xaxisspace)},anchor=north},
    y label style={at={(axis description cs:\yaxisspace,0.5)},anchor=south},
    xtick={1, 60, 600, 1800}, xticklabels={1, 60, 600, 1800},
    extra x ticks = {3600, 7200}, extra x tick labels = {OoR, Err}, extra x tick style = {grid = major},
    ytick={1, 60, 600, 1800}, yticklabels={1, 60, 600, 1800},
    extra y ticks = {3600, 7200}, extra y tick labels = {OoR, Err}, extra y tick style = {grid = major},
    xlabel=\storm (sparse), ylabel=\prism (explicit),
    yticklabel style={font=\tiny}, xticklabel style={rotate=290, anchor=west, font=\tiny},
    legend pos=south east, legend style={font=\tiny}]
  \addplot[
    scatter,only marks,
    scatter/classes={
      dtmcs={dtmc},
      mdps={mdp},
      ctmcs={ctmc}
    },
    scatter src=explicit symbolic]
    table [col sep=semicolon, x=Storm-sparse, y=Prism-sparse, meta=Type]
    {fig/plots/results.csv};
  \legend{}
  \addplot[no marks] coordinates
    {(0.01,0.01) (1800,1800) };
  \addplot[no marks, dashed] coordinates
    {(0.01,0.1) (180,1800) };
  \addplot[no marks, dashed] coordinates
    {(0.01,1) (18,1800) };
  \end{axis}
\end{tikzpicture}
    \end{minipage}
  \divider
    \begin{minipage}{0.28\textwidth}
      \centering
      \begin{tikzpicture}
  \begin{axis}[
    width=4.4cm, height=4.4cm, xmin=0.1, ymin=0.1, ymax=7200, xmax=7200, xmode=log, ymode=log,
    axis x line=bottom, axis y line=left,
    x label style={at={(axis description cs:0.5,\xaxisspace)},anchor=north},
    y label style={at={(axis description cs:\yaxisspace,0.5)},anchor=south},
    xtick={1, 60, 600, 1800}, xticklabels={1, 60, 600, 1800},
    extra x ticks = {3600, 7200}, extra x tick labels = {OoR, Err}, extra x tick style = {grid = major},
    ytick={1, 60, 600, 1800}, yticklabels={1, 60, 600, 1800},
    extra y ticks = {3600, 7200}, extra y tick labels = {OoR, Err}, extra y tick style = {grid = major},
    xlabel=\storm (hybrid), ylabel=\prism (sparse),
    yticklabel style={font=\tiny}, xticklabel style={rotate=290, anchor=west, font=\tiny},
    legend pos=south east, legend style={font=\tiny}]
  \addplot[
    scatter,only marks,
    scatter/classes={
      dtmcs={dtmc},
      mdps={mdp},
      ctmcs={ctmc}
    },
    scatter src=explicit symbolic]
    table [col sep=semicolon, x=Storm-hybrid, y=Prism-hybrid, meta=Type]
    {fig/plots/results.csv};
  \legend{}
  \addplot[no marks] coordinates
    {(0.01,0.01) (1800,1800) };
  \addplot[no marks, dashed] coordinates
    {(0.01,0.1) (180,1800) };
  \addplot[no marks, dashed] coordinates
    {(0.01,1) (18,1800) };
  \end{axis}
\end{tikzpicture}
    \end{minipage}
  \divider
    \begin{minipage}{0.28\textwidth}
      \centering
      \begin{tikzpicture}
  \begin{axis}[
    width=4.4cm, height=4.4cm, xmin=0.1, ymin=0.1, ymax=7200, xmax=7200, xmode=log, ymode=log,
    axis x line=bottom, axis y line=left,
    x label style={at={(axis description cs:0.5,\xaxisspace)},anchor=north},
    y label style={at={(axis description cs:\yaxisspace,0.5)},anchor=south},
    xtick={1, 60, 600, 1800}, xticklabels={1, 60, 600, 1800},
    extra x ticks = {3600, 7200}, extra x tick labels = {OoR, Err}, extra x tick style = {grid = major},
    ytick={1, 60, 600, 1800}, yticklabels={1, 60, 600, 1800},
    extra y ticks = {3600, 7200}, extra y tick labels = {OoR, Err}, extra y tick style = {grid = major},
    xlabel=\storm (dd), ylabel=\prism (mtbdd),
    yticklabel style={font=\tiny}, xticklabel style={rotate=290, anchor=west, font=\tiny},
    legend pos=south east, legend style={font=\tiny}]
  \addplot[
    scatter,only marks,
    scatter/classes={
      dtmcs={dtmc},
      mdps={mdp},
      ctmcs={ctmc}
    },
    scatter src=explicit symbolic]
    table [col sep=semicolon, x=Storm-cudd, y=Prism-cudd, meta=Type]
    {fig/plots/results.csv};
  \legend{}
  \addplot[no marks] coordinates
    {(0.01,0.01) (1800,1800) };
  \addplot[no marks, dashed] coordinates
    {(0.01,0.1) (180,1800) };
  \addplot[no marks, dashed] coordinates
    {(0.01,1) (18,1800) };
  \end{axis}
\end{tikzpicture}
    \end{minipage}
  \vspace{-4mm}
  \\
  
  \hspace{-0.46cm}
    \begin{minipage}{0.28\textwidth}
      \centering
      \begin{tikzpicture}
  \begin{axis}[
    width=4.4cm, height=4.4cm, xmin=0.1, ymin=0.1, ymax=3600, xmax=3600, xmode=log, ymode=log,
    axis x line=bottom, axis y line=left,
    x label style={at={(axis description cs:0.5,\xaxisspace)},anchor=north},
    y label style={at={(axis description cs:\yaxisspace,0.5)},anchor=south},
    xtick={1, 60, 600, 1800}, xticklabels={1, 60, 600, 1800},
    extra x ticks = {3600}, extra x tick labels = {NR}, extra x tick style = {grid = major},
    ytick={1, 60, 600, 1800}, yticklabels={1, 60, 600, 1800},
    extra y ticks = {3600}, extra y tick labels = {NR}, extra y tick style = {grid = major},
    xlabel=\storm (best), ylabel=\prism (best),
    yticklabel style={font=\tiny}, xticklabel style={rotate=290, anchor=west, font=\tiny},
    legend pos=south east, legend style={font=\tiny}]
  \addplot[
    scatter,only marks,
    scatter/classes={
      dtmcs={dtmc},
      mdps={mdp},
      ctmcs={ctmc}
    },
    scatter src=explicit symbolic]
    table [col sep=semicolon, x=Storm, y=Prism, meta=Type]
    {fig/plots/best_time.csv};
  \legend{}
  \addplot[no marks] coordinates
    {(0.01,0.01) (1800,1800) };
  \addplot[no marks, dashed] coordinates
    {(0.01,0.1) (180,1800) };
  \addplot[no marks, dashed] coordinates
    {(0.01,1) (18,1800) };
  \end{axis}
\end{tikzpicture}
    \end{minipage}
       \divider
    \begin{minipage}{0.28\textwidth}
      \centering
      \begin{tikzpicture}
  \begin{axis}[
    width=4.4cm, height=4.4cm, xmin=0.1, ymin=0.1, ymax=7200, xmax=7200, xmode=log, ymode=log,
    axis x line=bottom, axis y line=left,
    x label style={at={(axis description cs:0.5,\xaxisspace)},anchor=north},
    y label style={at={(axis description cs:\yaxisspace,0.5)},anchor=south},
    xtick={1, 60, 600, 1800}, xticklabels={1, 60, 600, 1800},
    extra x ticks = {3600, 7200}, extra x tick labels = {OoR, Err}, extra x tick style = {grid = major},
    ytick={1, 60, 600, 1800}, yticklabels={1, 60, 600, 1800},
    extra y ticks = {3600, 7200}, extra y tick labels = {OoR, Err}, extra y tick style = {grid = major},
    xlabel=\storm (exact), ylabel=\prism (exact),
    yticklabel style={font=\tiny}, xticklabel style={rotate=290, anchor=west, font=\tiny},
    legend pos=south east, legend style={font=\tiny}]
  \addplot[
    scatter,only marks,
    scatter/classes={
      dtmcs={dtmc},
      mdps={mdp},
      ctmcs={ctmc}
    },
    scatter src=explicit symbolic]
    table [col sep=semicolon, x=Storm-exact, y=Prism-exact, meta=Type]
    {fig/plots/results.csv};
  \legend{}
  \addplot[no marks] coordinates
    {(0.01,0.01) (1800,1800) };
  \addplot[no marks, dashed] coordinates
    {(0.01,0.1) (180,1800) };
  \addplot[no marks, dashed] coordinates
    {(0.01,1) (18,1800) };
  \end{axis}
\end{tikzpicture}
    \end{minipage}
  \divider
    \begin{minipage}{0.28\textwidth}
      \centering
      \begin{tikzpicture}
  \begin{axis}[
    width=4.4cm, height=4.4cm, xmin=0.1, ymin=0.1, ymax=7200, xmax=7200, xmode=log, ymode=log,
    axis x line=bottom, axis y line=left,
    x label style={at={(axis description cs:0.5,\xaxisspace)},anchor=north},
    y label style={at={(axis description cs:0.19,0.5)},anchor=south},
    xtick={1, 60, 600, 1800}, xticklabels={1, 60, 600, 1800},
    extra x ticks = {3600, 7200}, extra x tick labels = {OoR, Err}, extra x tick style = {grid = major},
    ytick={1, 60, 600, 1800}, yticklabels={1, 60, 600, 1800},
    extra y ticks = {3600, 7200}, extra y tick labels = {OoR, Err}, extra y tick style = {grid = major},
    xlabel=\storm (sparse), ylabel=\imca,
    yticklabel style={font=\tiny}, xticklabel style={rotate=290, anchor=west, font=\tiny},
    legend pos=south east, legend style={font=\tiny}]
  \addplot[
    scatter,only marks,
    scatter/classes={
      mas={ma}
    },
    scatter src=explicit symbolic]
    table [col sep=semicolon, x=Storm-ma, y=Imca-ma, meta=Type]
    {fig/plots/results_ma.csv};
  \legend{}
  \addplot[no marks] coordinates
    {(0.01,0.01) (1800,1800) };
  \addplot[no marks, dashed] coordinates
    {(0.01,0.1) (180,1800) };
  \addplot[no marks, dashed] coordinates
    {(0.01,1) (18,1800) };
  \end{axis}
\end{tikzpicture}
    \end{minipage}
  \\
  \vspace{-3mm}
  \centering
      \newenvironment{customlegend}[1][]{%
        \begingroup
        \csname pgfplots@init@cleared@structures\endcsname
        \pgfplotsset{#1}%
    }{%
        \csname pgfplots@createlegend\endcsname
        \endgroup
    }%
\def\addlegendimage{\csname pgfplots@addlegendimage\endcsname}

\begin{tikzpicture}
  \begin{customlegend}[legend columns=4,legend style={align=left,draw=none,column sep=2ex},legend entries={DTMC, MDP, CTMC, MA}]
    \addlegendimage{dtmc, only marks, mark size=2},
    \addlegendimage{mdp, only marks, mark size=2},
    \addlegendimage{ctmc, only marks, mark size=2}
    \addlegendimage{ma, only marks, mark size=2}
  \end{customlegend}
\end{tikzpicture}
  \vspace{-3mm}
  \caption{Run-time comparison (seconds) of different engines/features.}
  \label{Fig:ExpScatter}
  \vspace{-3mm}
\end{figure}
Fig.~\ref{Fig:ExpScatter} (top-row) summarizes the results of the experiments\cref{note:appendix} in log-log scale.
We plot the total time taken by the \storm engines versus the ``matching'' \prism engines.
Data points above the main diagonal indicate that \storm solved the task faster.
The two dashed lines indicate a speed-up of 10 and 100, respectively; ``OoR'' denotes memory- or time-outs, ``Err'' denotes that a tool was not able to complete the task for any other reason and ``NR'' stands for ``no result''.\cref{note:appendix}
\paragraph{Discussion.}
We observe that \storm is competitive on all compared engines.
Even though the MTBDD-based engines are very similar and even use the same MTBDD library (\tool{CUDD}), most of the time \storm is able to outperform \prism.
Note that \storm currently does not support CTMCs in this engine.
We observe a slightly clearer advantage of \storm' hybrid engine in comparison to \prism's sparse engine.
Here, model building times tend to be similar, but most often the numerical solution is done more efficiently by \storm.
However, for large CTMC benchmarks, \prism tends to be faster than \storm.
\storm's \engine{sparse} engine consistently outperforms \prism due to both the time needed for model construction as well as solving times.
For the overwhelming majority of verification tasks, \storm's best engine is faster than \prism's best engine. \storm solves 361 (out of 380) tasks, compared to 346 tasks \prism solves. 
\subsubsection{Exact arithmetic.}
Fig.~\ref{Fig:ExpScatter}(bottom center) compares the \emph{exact} modes of both tools.
\storm outperforms \prism by up to three orders of magnitude.
\subsubsection{Markov Automata.}
As \prism does not support the verification of MAs, we compare \storm with the only other tool capable of verifying MAs: \imca~\cite{DBLP:conf/atva/GuckTHRS14}.
We used the models provided by \imca, results are depicted in Fig.~\ref{Fig:ExpScatter}(bottom right).
For most instances, \storm is significantly faster.
  \small{
\subsubsection{Acknowledgments.}
\label{sec:Acknowledgments}
The authors would like to thank people that support(ed) the development of \storm over the years (in alphabetical order): Philipp Berger, Harold Bruintjes, Gereon Kremer, David Korzeniewski, and Tim Quatmann.}

  \bibliographystyle{splncs}
  \bibliography{literature}
  \clearpage
  \appendix
  \section{Appendix}
\newcommand{\plotsize}{6cm}
\newcommand{\yaxisappendix}{0.15}
\newcommand{\xaxisappendix}{-0.05}
\newcommand{\dividerappendix}{\hspace{0.1cm}}
\label{app:moreexperiments}
\subsection{Remarks}
\label{app:remarks}
\paragraph{More details on experiments.} Before we go into more detail and show experiments that were not included in the main paper, we want to point out that even more details on the conducted experiments can be found at 
\begin{center}
  \url{https://moves-rwth.github.io/storm/benchmarks.html}
\end{center}
\paragraph{Errors and unsupported tasks.} We decided to treat errors and unsupported tasks in the same way (by marking them as ``Err'').
The reason for this is that we do not see a fundamental difference between the two outcomes.
Every tool could in principle catch errors and report that the task is not supported.
Similarly, unsupported tasks could potentially lead to other errors if inputs are not sufficiently checked.
Finally, from a user perspective, both outcomes mean that no answer is available. 

In the plots that compare the best run-times for each individual experiment, it is unclear how to label an instance if no engine could solve the task, as there is no natural ordering on error (``Err'') and time- or memory-outs (``OoR'').
We therefore resort to label these points with \emph{no result} (``NR'').
\paragraph{Memory consumption.} In our experiments, we only compared the run-times of the tools on the benchmark instances.
Obviously, this is not the only interesting metric as (probabilistic) model checking is a memory-intensive task.
Verification might fail because of insufficient memory and the memory footprints of the tools may govern whether a model can be treated or not.
However, in this particular comparison, this raises several issues. 
First, comparing memory consumption of Java and \texttt{C++} programs is hard as the Java VM (unlike \texttt{C++} applications) only releases memory when it needs to and therefore will approach the given memory limit (for larger benchmarks).
Nevertheless, this does not imply that this memory was necessary as earlier garbage collection might have released enough memory to stay below some bound.
Second, measuring the actual size in memory is a notoriously hard problem and virtual memory consumption might not properly reflect the actual memory consumption.
\subsection{\iscasmc vs. \storm}
We did not include the comparison with \iscasmc in the main paper as the two tools clearly have different objectives: \iscasmc focusses on complex linear time properties whereas \storm focuses on branching time logics.
However, thanks to the authors of \iscasmc, we were able to obtain a snapshot of \epmc, the upcoming successor of \iscasmc.
\epmc has not been developed from scratch, but rather builds on the basis of \iscasmc.
According to the authors, major refactorings were conducted.
We compare the tools' performances on the \prism benchmark suite.
Again, we compare ``matching'' engines: \storm's \engine{sparse} engine against \epmc's \engine{sparse} and \storm's \engine{dd} engine versus \epmc's \engine{dd} engine.
As it is currently not clear to us, whether \epmc's \engine{dd} engine is fully symbolic as \storm's \engine{dd} engine or rather follows a hybrid approach, we also include the comparison of this engine with storm's \engine{hybrid} engine.
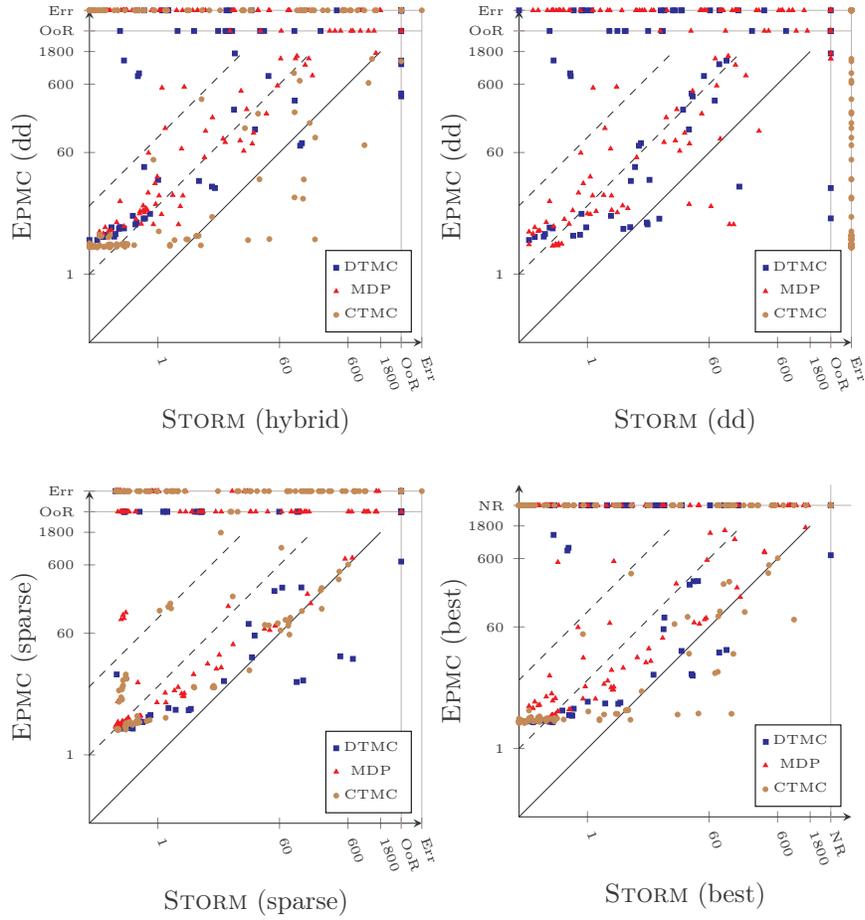
\begin{figure}[ht]
  \centering
  \begin{minipage}{0.45\textwidth}
    \begin{tikzpicture}
  \begin{axis}[
    width=\plotsize, height=\plotsize, xmin=0.1, ymin=0.1, ymax=7200, xmax=7200, xmode=log, ymode=log,
    axis x line=bottom, axis y line=left,
    x label style={at={(axis description cs:0.5,\xaxisappendix)},anchor=north},
    y label style={at={(axis description cs:\yaxisappendix,0.5)},anchor=south},
    xtick={1, 60, 600, 1800}, xticklabels={1, 60, 600, 1800},
    extra x ticks = {3600, 7200}, extra x tick labels = {OoR, Err}, extra x tick style = {grid = major},
    ytick={1, 60, 600, 1800}, yticklabels={1, 60, 600, 1800},
    extra y ticks = {3600, 7200}, extra y tick labels = {OoR, Err}, extra y tick style = {grid = major},
    xlabel=\storm (hybrid), ylabel=\epmc (dd),
    yticklabel style={font=\tiny}, xticklabel style={rotate=290, anchor=west, font=\tiny},
    legend pos=south east, legend style={font=\tiny}]
  \addplot[
    scatter,only marks,
    scatter/classes={
      dtmcs={dtmc},
      mdps={mdp},
      ctmcs={ctmc}
    },
    scatter src=explicit symbolic]
    table [col sep=semicolon, x=Storm-hybrid, y=Epmc-cudd, meta=Type]
    {fig/plots/results.csv};
  \legend{DTMC,MDP,CTMC} 
  \addplot[no marks] coordinates
    {(0.01,0.01) (1800,1800) };
  \addplot[no marks, dashed] coordinates
    {(0.01,0.1) (180,1800) };
  \addplot[no marks, dashed] coordinates
    {(0.01,1) (18,1800) };
  \end{axis}
\end{tikzpicture}
  \end{minipage}
  \dividerappendix
  \begin{minipage}{0.45\textwidth}
    \begin{tikzpicture}
  \begin{axis}[
    width=\plotsize, height=\plotsize, xmin=0.1, ymin=0.1, ymax=7200, xmax=7200, xmode=log, ymode=log,
    axis x line=bottom, axis y line=left,
    x label style={at={(axis description cs:0.5,\xaxisappendix)},anchor=north},
    y label style={at={(axis description cs:\yaxisappendix,0.5)},anchor=south},
    xtick={1, 60, 600, 1800}, xticklabels={1, 60, 600, 1800},
    extra x ticks = {3600, 7200}, extra x tick labels = {OoR, Err}, extra x tick style = {grid = major},
    ytick={1, 60, 600, 1800}, yticklabels={1, 60, 600, 1800},
    extra y ticks = {3600, 7200}, extra y tick labels = {OoR, Err}, extra y tick style = {grid = major},
    xlabel=\storm (dd), ylabel=\epmc (dd),
    yticklabel style={font=\tiny}, xticklabel style={rotate=290, anchor=west, font=\tiny},
    legend pos=south east, legend style={font=\tiny}]
  \addplot[
    scatter,only marks,
    scatter/classes={
      dtmcs={dtmc},
      mdps={mdp},
      ctmcs={ctmc}
    },
    scatter src=explicit symbolic]
    table [col sep=semicolon, x=Storm-cudd, y=Epmc-cudd, meta=Type]
    {fig/plots/results.csv};
  \legend{DTMC,MDP,CTMC} 
  \addplot[no marks] coordinates
    {(0.01,0.01) (1800,1800) };
  \addplot[no marks, dashed] coordinates
    {(0.01,0.1) (180,1800) };
  \addplot[no marks, dashed] coordinates
    {(0.01,1) (18,1800) };
  \end{axis}
\end{tikzpicture}
  \end{minipage}
  \dividerappendix
  \begin{minipage}{0.45\textwidth}
    \begin{tikzpicture}
  \begin{axis}[
    width=\plotsize, height=\plotsize, xmin=0.1, ymin=0.1, ymax=7200, xmax=7200, xmode=log, ymode=log,
    axis x line=bottom, axis y line=left,
    x label style={at={(axis description cs:0.5,\xaxisappendix)},anchor=north},
    y label style={at={(axis description cs:\yaxisappendix,0.5)},anchor=south},
    xtick={1, 60, 600, 1800}, xticklabels={1, 60, 600, 1800},
    extra x ticks = {3600, 7200}, extra x tick labels = {OoR, Err}, extra x tick style = {grid = major},
    ytick={1, 60, 600, 1800}, yticklabels={1, 60, 600, 1800},
    extra y ticks = {3600, 7200}, extra y tick labels = {OoR, Err}, extra y tick style = {grid = major},
    xlabel=\storm (sparse), ylabel=\epmc (sparse),
    yticklabel style={font=\tiny}, xticklabel style={rotate=290, anchor=west, font=\tiny},
    legend pos=south east, legend style={font=\tiny}]
  \addplot[
    scatter,only marks,
    scatter/classes={
      dtmcs={dtmc},
      mdps={mdp},
      ctmcs={ctmc}
    },
    scatter src=explicit symbolic]
    table [col sep=semicolon, x=Storm-sparse, y=Epmc-sparse, meta=Type]
    {fig/plots/results.csv};
  \legend{DTMC,MDP,CTMC} 
  \addplot[no marks] coordinates
    {(0.01,0.01) (1800,1800) };
  \addplot[no marks, dashed] coordinates
    {(0.01,0.1) (180,1800) };
  \addplot[no marks, dashed] coordinates
    {(0.01,1) (18,1800) };
  \end{axis}
\end{tikzpicture}
  \end{minipage}
  \dividerappendix
  \begin{minipage}{0.45\textwidth}
    \begin{tikzpicture}
  \begin{axis}[
    width=\plotsize, height=\plotsize, xmin=0.1, ymin=0.1, ymax=7200, xmax=7200, xmode=log, ymode=log,
    axis x line=bottom, axis y line=left,
    x label style={at={(axis description cs:0.5,\xaxisappendix)},anchor=north},
    y label style={at={(axis description cs:\yaxisappendix,0.5)},anchor=south},
    xtick={1, 60, 600, 1800}, xticklabels={1, 60, 600, 1800},
    extra x ticks = {3600}, extra x tick labels = {NR}, extra x tick style = {grid = major},
    ytick={1, 60, 600, 1800}, yticklabels={1, 60, 600, 1800},
    extra y ticks = {3600}, extra y tick labels = {NR}, extra y tick style = {grid = major},
    xlabel=\storm (best), ylabel=\epmc (best),
    yticklabel style={font=\tiny}, xticklabel style={rotate=290, anchor=west, font=\tiny},
    legend pos=south east, legend style={font=\tiny}]
  \addplot[
    scatter,only marks,
    scatter/classes={
      dtmcs={dtmc},
      mdps={mdp},
      ctmcs={ctmc}
    },
    scatter src=explicit symbolic]
    table [col sep=semicolon, x=Storm, y=Epmc, meta=Type]
    {fig/plots/best_time.csv};
  \legend{DTMC,MDP,CTMC} 
  \addplot[no marks] coordinates
    {(0.01,0.01) (1800,1800) };
  \addplot[no marks, dashed] coordinates
    {(0.01,0.1) (180,1800) };
  \addplot[no marks, dashed] coordinates
    {(0.01,1) (18,1800) };
  \end{axis}
\end{tikzpicture}
  \end{minipage}
  \caption{A comparison of \epmc and \storm.}
  \label{fig:storm_vs_epmc}
\end{figure}
Figure \ref{fig:storm_vs_epmc} shows the plots obtained from the experiments.
Note that \epmc does not yet support expected reward objectives and steady-state objectives, which explains (most of) the data points labeled with ``Err''.
\subsection{Accumulated time plots (\prism vs. \storm vs. \epmc)}
\begin{figure}[ht]
  \centering
  \begin{tikzpicture}
  \begin{axis}[
    width=\plotsize, height=\plotsize, xmin=0, ymin=1, xmax=380, ymax=50000, ymode=log,
    axis x line=bottom, axis y line=left,
    x label style={at={(axis description cs:0.5,\xaxisappendix)},anchor=north},
    y label style={at={(axis description cs:\yaxisappendix,0.5)},anchor=south},
    xlabel=No. benchmark, ylabel=Time (in seconds),
    yticklabel style={font=\tiny}, xticklabel style={rotate=290, anchor=west, font=\tiny},
    mark repeat={50},
    legend pos=south east, legend style={font=\tiny}]
  \addplot[color=orange,mark=x]
    table [col sep=semicolon, x=no, y=Storm]
    {fig/plots/best_cumulative_plot.csv};
  \addplot[color=blue,mark=*]
    table [col sep=semicolon, x=no, y=Prism]
    {fig/plots/best_cumulative_plot.csv};
  \addplot[color=red,mark=square*]
    table [col sep=semicolon, x=no, y=Epmc]
    {fig/plots/best_cumulative_plot.csv};
  \legend{\storm,\prism,\epmc}
  \end{axis}
\end{tikzpicture}
  \caption{\prism vs. \epmc vs. \storm (best engine for each instance)}
  \label{fig:storm_vs_prism_svcomp}
\end{figure}
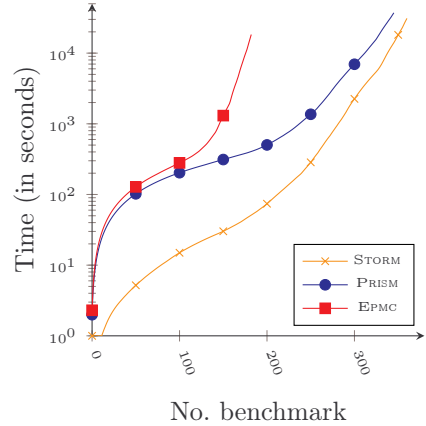
Figure \ref{Fig:ExpScatter} (bottom left) in the paper shows a plot of \storm's and \prism's running times where for each individual instance we picked the running time of the fastest engine (for that particular instance).
Figure \ref{fig:storm_vs_prism_svcomp} shows the same data, but in a different format.
In the spirit of the ``score-based quantile plots'' used by the well-known competition on software verification (SVCOMP), we give the accumulated time (again taking the fastest engine for each particular instance) over the number of instances solved.
That is, a point at $(x, y)$ represents that the fastest $x$ instances were solved in accumulated time $y$.
\subsection{\prism's ``hybrid'' engine}
\begin{figure}[ht]
  \centering
  \begin{tikzpicture}
  \begin{axis}[
    width=\plotsize, height=\plotsize, xmin=0.1, ymin=0.1, ymax=7200, xmax=7200, xmode=log, ymode=log,
    axis x line=bottom, axis y line=left,
    x label style={at={(axis description cs:0.5,\xaxisappendix)},anchor=north},
    y label style={at={(axis description cs:\yaxisappendix,0.5)},anchor=south},
    xtick={1, 60, 600, 1800}, xticklabels={1, 60, 600, 1800},
    extra x ticks = {3600, 7200}, extra x tick labels = {OoR, Err}, extra x tick style = {grid = major},
    ytick={1, 60, 600, 1800}, yticklabels={1, 60, 600, 1800},
    extra y ticks = {3600, 7200}, extra y tick labels = {OoR, Err}, extra y tick style = {grid = major},
    xlabel=\prism (hybrid), ylabel=\prism (sparse),
    yticklabel style={font=\tiny}, xticklabel style={rotate=290, anchor=west, font=\tiny},
    legend pos=south east, legend style={font=\tiny}]
  \addplot[
    scatter,only marks,
    scatter/classes={
      dtmcs={dtmc},
      mdps={mdp},
      ctmcs={ctmc}
    },
    scatter src=explicit symbolic]
    table [col sep=semicolon, x=Prism-own-hybrid, y=Prism-hybrid, meta=Type]
    {fig/plots/results.csv};
  \legend{DTMC,MDP,CTMC} 
  \addplot[no marks] coordinates
    {(0.01,0.01) (1800,1800) };
  \addplot[no marks, dashed] coordinates
    {(0.01,0.1) (180,1800) };
  \addplot[no marks, dashed] coordinates
    {(0.01,1) (18,1800) };
  \end{axis}
\end{tikzpicture}
  \caption{A comparison of \prism's \engine{mtbdd} and \engine{sparse} engines.}
  \label{fig:prism_hybrid_vs_sparse}
\end{figure}
\storm does currently not feature an engine that is comparable to \prism's ``hybrid'' engine.
The latter is a sophisticated cross-over between its \engine{sparse} and \engine{mtbdd} engines.
While this engine is optimized towards a good space-time trade-off, our main objective is performance in terms of time.
To illustrate that \prism's \engine{hybrid} engine is dominated (in terms of run-time) by \prism's own \engine{sparse} engine, consider Figure \ref{fig:prism_hybrid_vs_sparse}.
Therefore, if running-times are the only objective, there is little motivation to also include an engine similar to \prism's \engine{hybrid} engine. Also, it justifies that we did not include \prism's \engine{hybrid} engine in the comparisons in the main paper.
\subsection{MTBDD-based CTMC model checking}
\begin{figure}[ht]
  \centering
  \begin{tikzpicture}
  \begin{axis}[
    width=\plotsize, height=\plotsize, xmin=0.1, ymin=0.1, ymax=7200, xmax=7200, xmode=log, ymode=log,
    axis x line=bottom, axis y line=left,
    x label style={at={(axis description cs:0.5,\xaxisappendix)},anchor=north},
    y label style={at={(axis description cs:\yaxisappendix,0.5)},anchor=south},
    xtick={1, 60, 600, 1800}, xticklabels={1, 60, 600, 1800},
    extra x ticks = {3600, 7200}, extra x tick labels = {OoR, Err}, extra x tick style = {grid = major},
    ytick={1, 60, 600, 1800}, yticklabels={1, 60, 600, 1800},
    extra y ticks = {3600, 7200}, extra y tick labels = {OoR, Err}, extra y tick style = {grid = major},
    xlabel=\prism (sparse), ylabel=\prism (dd),
    yticklabel style={font=\tiny}, xticklabel style={rotate=290, anchor=west, font=\tiny},
    legend pos=south east, legend style={font=\tiny}]
  \addplot[
    scatter,only marks,
    scatter/classes={
      dtmcs={dtmc},
      mdps={mdp},
      ctmcs={ctmc}
    },
    scatter src=explicit symbolic]
    table [col sep=semicolon, x=Prism-hybrid, y=Prism-cudd, meta=Type]
    {fig/plots/results.csv};
  \legend{DTMC,MDP,CTMC} 
  \addplot[no marks] coordinates
    {(0.01,0.01) (1800,1800) };
  \addplot[no marks, dashed] coordinates
    {(0.01,0.1) (180,1800) };
  \addplot[no marks, dashed] coordinates
    {(0.01,1) (18,1800) };
  \end{axis}
\end{tikzpicture}
  \caption{A comparison of \prism's \engine{mtbdd} and \engine{sparse} engines.}
  \label{fig:ctmc_dd}
\end{figure}
\storm does currently not feature fully symbolic (MTBDD-based) CTMC model checking.
To justify this, consider the plot in Figure \ref{fig:ctmc_dd}.
It compares the performance of two of \prism's engines, namely the fully symbolic one (\engine{mtbdd}) and \engine{sparse} on the \prism benchmark suite.
It can be easily observed that for CTMCs, the \engine{sparse} engine typically beats the MTBDD-based model checking in terms of time.
We are, however, well aware that for example memory consumption is potentially lower in the \engine{mtbdd} engine. As argued above (cf. remarks in Appendix \ref{app:remarks}), we consider run-times only.
\subsection{\storm vs. \storm}
Finally, we want to illustrate that within \storm, no engine clearly beats the others.
To this end, we also compare every of \storm's engines with every other.
Figure \ref{fig:storm_vs_storm} shows the results. While the \engine{hybrid} engine tends to dominate the \engine{dd} engine, there are several large examples that the latter can handle on which the former runs out of resources.
\begin{figure}[ht]
  \centering
  \begin{minipage}{0.45\textwidth}
    \begin{tikzpicture}
  \begin{axis}[
    width=\plotsize, height=\plotsize, xmin=0.1, ymin=0.1, ymax=7200, xmax=7200, xmode=log, ymode=log,
    axis x line=bottom, axis y line=left,
    x label style={at={(axis description cs:0.5,\xaxisappendix)},anchor=north},
    y label style={at={(axis description cs:\yaxisappendix,0.5)},anchor=south},
    xtick={1, 60, 600, 1800}, xticklabels={1, 60, 600, 1800},
    extra x ticks = {3600, 7200}, extra x tick labels = {OoR, Err}, extra x tick style = {grid = major},
    ytick={1, 60, 600, 1800}, yticklabels={1, 60, 600, 1800},
    extra y ticks = {3600, 7200}, extra y tick labels = {OoR, Err}, extra y tick style = {grid = major},
    xlabel=\storm (sparse), ylabel=\storm (dd),
    yticklabel style={font=\tiny}, xticklabel style={rotate=290, anchor=west, font=\tiny},
    legend pos=south east, legend style={font=\tiny}]
  \addplot[
    scatter,only marks,
    scatter/classes={
      dtmcs={dtmc},
      mdps={mdp},
      ctmcs={ctmc}
    },
    scatter src=explicit symbolic]
    table [col sep=semicolon, x=Storm-sparse, y=Storm-cudd, meta=Type]
    {fig/plots/results.csv};
  \legend{DTMC,MDP,CTMC} 
  \addplot[no marks] coordinates
    {(0.01,0.01) (1800,1800) };
  \addplot[no marks, dashed] coordinates
    {(0.01,0.1) (180,1800) };
  \addplot[no marks, dashed] coordinates
    {(0.01,1) (18,1800) };
  \end{axis}
\end{tikzpicture}
  \end{minipage}
  \dividerappendix
  \begin{minipage}{0.45\textwidth}
    \begin{tikzpicture}
  \begin{axis}[
    width=\plotsize, height=\plotsize, xmin=0.1, ymin=0.1, ymax=7200, xmax=7200, xmode=log, ymode=log,
    axis x line=bottom, axis y line=left,
    x label style={at={(axis description cs:0.5,\xaxisappendix)},anchor=north},
    y label style={at={(axis description cs:\yaxisappendix,0.5)},anchor=south},
    xtick={1, 60, 600, 1800}, xticklabels={1, 60, 600, 1800},
    extra x ticks = {3600, 7200}, extra x tick labels = {OoR, Err}, extra x tick style = {grid = major},
    ytick={1, 60, 600, 1800}, yticklabels={1, 60, 600, 1800},
    extra y ticks = {3600, 7200}, extra y tick labels = {OoR, Err}, extra y tick style = {grid = major},
    xlabel=\storm (sparse), ylabel=\storm (hybrid),
    yticklabel style={font=\tiny}, xticklabel style={rotate=290, anchor=west, font=\tiny},
    legend pos=south east, legend style={font=\tiny}]
  \addplot[
    scatter,only marks,
    scatter/classes={
      dtmcs={dtmc},
      mdps={mdp},
      ctmcs={ctmc}
    },
    scatter src=explicit symbolic]
    table [col sep=semicolon, x=Storm-sparse, y=Storm-hybrid, meta=Type]
    {fig/plots/results.csv};
  \legend{DTMC,MDP,CTMC} 
  \addplot[no marks] coordinates
    {(0.01,0.01) (1800,1800) };
  \addplot[no marks, dashed] coordinates
    {(0.01,0.1) (180,1800) };
  \addplot[no marks, dashed] coordinates
    {(0.01,1) (18,1800) };
  \end{axis}
\end{tikzpicture}
   \end{minipage}
  \dividerappendix
  \begin{minipage}{0.45\textwidth}
    \begin{tikzpicture}
  \begin{axis}[
    width=\plotsize, height=\plotsize, xmin=0.1, ymin=0.1, ymax=7200, xmax=7200, xmode=log, ymode=log,
    axis x line=bottom, axis y line=left,
    x label style={at={(axis description cs:0.5,\xaxisappendix)},anchor=north},
    y label style={at={(axis description cs:\yaxisappendix,0.5)},anchor=south},
    xtick={1, 60, 600, 1800}, xticklabels={1, 60, 600, 1800},
    extra x ticks = {3600, 7200}, extra x tick labels = {OoR, Err}, extra x tick style = {grid = major},
    ytick={1, 60, 600, 1800}, yticklabels={1, 60, 600, 1800},
    extra y ticks = {3600, 7200}, extra y tick labels = {OoR, Err}, extra y tick style = {grid = major},
    xlabel=\storm (dd), ylabel=\storm (hybrid),
    yticklabel style={font=\tiny}, xticklabel style={rotate=290, anchor=west, font=\tiny},
    legend pos=south east, legend style={font=\tiny}]
  \addplot[
    scatter,only marks,
    scatter/classes={
      dtmcs={dtmc},
      mdps={mdp},
      ctmcs={ctmc}
    },
    scatter src=explicit symbolic]
    table [col sep=semicolon, x=Storm-cudd, y=Storm-hybrid, meta=Type]
    {fig/plots/results.csv};
  \legend{DTMC,MDP,CTMC} 
  \addplot[no marks] coordinates
    {(0.01,0.01) (1800,1800) };
  \addplot[no marks, dashed] coordinates
    {(0.01,0.1) (180,1800) };
  \addplot[no marks, dashed] coordinates
    {(0.01,1) (18,1800) };
  \end{axis}
\end{tikzpicture}
  \end{minipage}
  \caption{A comparison of \storm's \engine{mtbdd} and \engine{sparse} engines.}
  \label{fig:storm_vs_storm}
\end{figure}

\end{document}